\documentstyle[12pt]{article}

\begin{document}

\title{A family of quantum interiors for ordinary and stringy black holes\,
II: Geometric considerations and global properties}

\author{Emilio Elizalde$^{\cite{email1,http}}$ and
Sergi R. Hildebrandt$^{\cite{email2}}$\\
Instituto de Ciencias del Espacio (CSIC) \& \\
Institut d'Estudis Espacials de Catalunya (IEEC/CSIC) \\
Edifici Nexus, Gran Capit\`a 2-4, 08034 Barcelona, Spain}
\date{}
\maketitle

% DEFINITIONS

\newcommand{\rf}[1]{{\rm (\ref{#1})}}

\def\ab{\alpha \beta}
\def\lh{\hbox{{\boldmath \hbox{$ \ell $}}} }
\def\th{\hbox{{\boldmath \hbox{$ \Theta $}}} }      % cobasis, Theta %
\def\l{\Lambda}
\def\le{\Lambda_1} % Lambda, exterior %
\def\li{\Lambda_2} % Lambda, interior $
\def\tr{{\tilde r} }
\def\dal{\sqcap_{\smash{\hskip -5.7pt \lower 1.3pt \hbox{--}}}\,\! }

\def\bce{\begin{center}}
\def\ece{\end{center}}
\def\beq{\begin{eqnarray}}
\def\eeq{\end{eqnarray}}
\def\ben{\begin{enumerate}}
\def\een{\end{enumerate}}
\def\ul{\underline}
\def\ni{\noindent}
\def\nn{\nonumber}
\def\bs{\bigskip}
\def\ms{\medskip}

% TEXT
\begin{abstract}
A family of spacetimes suitable for describing the matter conditions of a
static, spherically symmetric quantum vacuum is studied, as well as its
reliability for describing a regular model for the interior of a
semiclassical, static black hole ---without ever invoking a mass shell for
the final object. In paper I, this condition was seen to limit the search
to only one, distinguished family, that was investigated in detail.
Here it will be proven that, aside from being mathematically generic
(in its uniqueness), this family exhibits beautiful
physical properties, that one would reasonably demand from a collapse 
process,
including the remarkable result  that isotropization may take place
conveniently far from the (unavoidable) regularization scale.
The analysis is also extended in order
to include the possibility of a stringy core, always within the limits 
imposed
by the semiclassical approach to gravitation. This constitutes a first
approximation to the final goal of trying to characterize a
regular,  self-gravitating, stringy black hole.
\end{abstract}

\newpage

\section{Introduction}
This is the second of a couple of works devoted to the study of the
spacetime properties of regular interiors for non-rotating black holes. In
the first one \cite{rqibh} (referred to as paper I in the sequel), we
identified and studied in detail a generic set of spacetimes, called
GNRSS spacetimes. We also recovered
previous attempts to understand this issue,  and extended them,
including a starting protocol for the quantization of the sources that
may generate these effects. However, a proof of their uniqueness for
describing the physics involved was still lacking.
It will be proven here that,
besides having a  beautiful characterization
from the mathematical point of view (concerning, in particular, its
universality and uniqueness,
under generic conditions of clear physical meaning),
the GNRSS spacetimes do fulfill all the
energy-stress properties currently expected to be associated to quantum
effects in strong fields affecting
non-rotating collapsed bodies.

The present work is twofold. In the first part,
after reviewing, in Sect.~\ref{s-fdssv},  the construction of
the generic families of static, spherically symmetric, quantum vacuums 
(SSQV),
in Sect.~\ref{s-gpssqv} the geometrical properties of the
solutions will be investigated, as well as those corresponding to
the isotropic vacuums, which might describe the
core of the object owing to the dominance of vacuum polarization.
In Sect.~\ref{s-srbh}, we will deal with the delicate question of
matching those interiors with an exterior black hole metric, without having
to invoke singular massive shells or the like. We will see
how this condition limits the search to only
one family, which will be mathematically characterized in full, and that it
turns out to be physically suitable for all our
aims. In Sect.~\ref{ss-cdsg}, we solve
a model for a regular black hole that does not assume
isotropization to occur at the origin.
Eventually, we recover the results of I, also in this case,
thus completing the scheme started there.

On the other hand, in Sect.~\ref{s-eip}, we turn to the issue of
replacing the interior, or at least the core of the collapsed object by
a stringy black hole. To that end, in Sect.~\ref{s-mssss},
we write the junction conditions for
two spherically symmetric spacetimes. In Sect.~\ref{s-assbh},
we apply them to a
supersymmetric stringy black hole,
always within the limits imposed by the
semiclassical approach to gravitation,
in order to obtain, at least, a first order approximation to
our ultimate goal, which is
to deal with a (regular) self-gravitating stringy black hole.
The paper ends with some conclusions.

\section{The families describing spherically symmetric \allowbreak
vacuums}
\label{s-fdssv}
If $ T_{\ab} $ is the stress-momentum tensor of a physical system,
the classical vacuum is defined by $ T_{\ab} = 0 $. Quantum vacuums
are usually defined through a cosmological constant, $ \Lambda $, e.g.
$ T_{\ab} = \Lambda g_{\ab} $, which yields, e.g., the de Sitter spacetimes.
Because of  various physical arguments (see e.g. \cite{glinner}--\cite{cvs})
it is expected that these
spacetimes describe, in first approximation, the core of a collapsed
object, i.e. a black hole. Nevertheless, a de Sitter spacetime cannot be
matched with a black hole exterior solution. A straightforward
  way out consists in accepting
the presence of singular mass shells \cite{ip,fmm,bp}, but this introduces
too high arbitrariness, as mentioned elsewhere. A different solution
is to extend the definition of quantum vacuum.
For the problem we want to deal with, which exhibits spherical symmetry,
one may define that a solution of Einstein's equations corresponds to a SSQV
whenever $ T_{\ab} \neq 0 $
and $ T^0_0 = T^1_1 $, $ T^2_2 = T^3_3 $, for some orthonormal
cobasis ---that of a local observer. In this way, any local observer
adapted to the
spherical symmetry will measure exactly the same values of $ T_{\ab} $ (see
e.g. \cite{dymni,zn}). For the moment, we shall only consider static 
vacuums,
because the first aim will be to describe the spacetime structure of
the interior of a regular black hole.

We shall now give the families of spacetimes that are suitable to
become SSQVs.
Any static spherically symmetric spacetime can be conveniently
described by
\beq
\label{met1}
ds^2 = -F(r) \, dt^2+ F^{-1}(r) \, dr^2 + G^2(r) \, d\Omega^2,
\eeq
where $ d\Omega^2 \equiv d\theta^2 + \sin^2\theta \, d\varphi^2 $.
There
are certainly other ways to represent these spacetimes, which avoid the
problems occurring near the
possible horizons, or by putting $ R^2 d\Omega^2 $, provided
$ G' \equiv  dG(r)/dr \neq 0 $ (see e.g. \cite{msi,wheeler}). Later,
we will deal with some of these possibilities, but here it will
suffice to consider the former representation.
A standard calculation of $ T_{\ab} $ yields, for a local
observer at rest with respect to the coordinate grid of~\rf{met1},
\beq
\rho = {1 \over G^2} \Bigl[ 1 - F(G'^2 + 2 G G'') - G G' F' \Bigr], \\
\label{eq-p}
p = {1 \over G^2} \Bigl[ -1 + F G'^2 + G G' F' \Bigr], \\
p_2 = p_3 = {F'' \over 2} + {F G'' \over G} + {F' G' \over G}.
\eeq
Imposing $ T^0_0 = T^1_1 $, $ T^2_2 = T^3_3 $ requires studying
the condition $ \rho + p = 0 $. This yields (for $ G \neq 0 $)
\beq
F \, G'' = 0.
\eeq
If $ F = 0 $, the expression~\rf{met1} is useless. In this case
the spacetimes can be written as\footnote{It is first necessary to change
the coordinate system of~\rf{met1} by $ dT \equiv dt + (1-F)/F dr $
while keeping the rest unchanged. Then one can impose $ F = 0 $.}
\beq
\label{eq-fn0}
ds^2 = 2 \, dT \, dr + 2 \,dr^2 + G^2(r) \, d\Omega^2.
\eeq
In the orthonormalized cobasis given by
\beq
\th^0 = dT/\sqrt{2}, \quad \th^1 = dT/\sqrt{2} + \sqrt{2} \, dr, \quad \th^2 
= G
\, d\theta, \quad \th^3 = G \sin\theta \, d\varphi,
\eeq
the Ricci tensor takes the form
\beq
\begin{array}{ll}
\label{ricci0}
\hbox{\bf Ricci} = & {\displaystyle {G'' \over G}} ( - \th^0 \otimes \th^0 +
\th^0 \otimes \th^1 + \th^1 \otimes \th^0 - \th^1 \otimes \th^1) \cr
& + \ {\displaystyle{1 \over G^2}} ( \th^2 \otimes \th^2 + \th^3 \otimes 
\th^3).
\end{array}
\eeq
While, for a SSQV we must have (the conditions on $ T_{\ab} $ being directly
translated into conditions for $ R_{\ab} $)
\beq
\label{ricci}
\hbox{\bf Ricci} = R_{00}( - \th^0_N \otimes \th^0_N + \th^1_N \otimes 
\th^1_N)
+ R_{22} ( \th^2_N \otimes \th^2_N + \th^3_N \otimes \th^3_N),
\eeq
where $ \{ \th^{\Omega}_N \} $ is some orthonormalized cobasis,
not necessarily coincident with the one used in the computation 
of~\rf{ricci0}.
Therefore, we must look for an orthonormalized cobasis for which the Ricci
tensor~\rf{ricci0} becomes of the type~\rf{ricci}. Clearly this is the same 
as
finding out whether we can have linear expressions
$ \th^0_N \equiv A \th^0 + B \th^1 $, and
$ \th^1_N = C \th^0 + D \th^1 $, with
$ -\th^0_N \cdot \th^0_N =  \th^1_N \cdot \th^1_N =
\th^0_N \cdot \th^1_N + 1= 1 $. However, the form
of the SSQV is invariant under these changes (for
they are adapted, by definition, to the spherical symmetry).
The only solution is
\beq
G'' = 0.
\eeq
If $ F \neq 0 $, we also have $ G'' = 0 $. Thus $ G'' = 0 $ constitutes the
proper characterization of any possibility.

From $ G'' = 0 $ two distinct alternatives appear
\beq
G = \gamma , \quad \hbox{or} \quad G= \alpha \, r + \gamma,
\eeq
where $ \alpha (\neq 0) $ and $ \gamma $ are constant.
In order to include the possible horizons, we write the
metrics~\rf{met1} under the common form
\beq
\label{met2}
ds^2 = - (1-H) \, dT^2 + 2 H \, dT \, dr + ( 1 + H ) \, dr^2
+ \gamma^2 \, d\Omega^2,
\\
\label{met-3}
ds^2 = - (1-H) \, dT^2 + 2 H \, dT \, dr + ( 1 + H ) \, dr^2
+ (\alpha r + \gamma)^2 \, d\Omega^2,
\eeq
where $ H \equiv 1 - F $, and the coordinate change has been
given elsewhere. The second case is in fact equivalent to
the case $ \alpha =1 $, $ \gamma = 0 $, as we shall now prove,
because we are dealing with families of spacetimes.

If we perform the coordinate change $ dT\! = \!\alpha d{\tilde t}
+ (\alpha^2-1) dr $, $ {\tilde r}\! \equiv \!\alpha r + \gamma $, leaving
$ \theta $ and $ \varphi $ unchanged, we get
\beq
\begin{array}{rl}
ds^2 = &- \alpha^2(1-H) \, d{\tilde t}^2
+ 2 (\alpha^2 H +1 -\alpha^2) \, d{\tilde t} \, d{\tilde r}
+ ( 2- \alpha^2 + \alpha^2 H) \, d{\tilde r}^2 + \nn \\
& {\tilde r}^2 \, d \Omega^2 \cr
\end{array}
\eeq
which shows that, by choosing
$ {\tilde H} \equiv \alpha^2 H + 1 - \alpha^2 $, one has
\beq
\label{met-gnrss}
ds^2 = -(1-{\tilde H}) \, d{\tilde t}^2 + 2 {\tilde H} \,
d{\tilde t} \, d{\tilde r} + (1 + {\tilde H}) \, d{\tilde r}^2 + {\tilde 
r}^2 \,
d\Omega^2.
\eeq

To summarize, there are {\it only} two ---non-equivalent--- families
of SSQV. The case with $ G' = 0 $ is characteristic of the Nariai
solution \cite{nariai,kramer}. The Nariai solution is a solution of
Einstein's equations for the same pattern as the de Sitter
solution, i.e. $ T_{\ab} = \Lambda_0 g_{\ab} $, being $ \Lambda_0 $
the cosmological constant. The difference lies in the ``radial'' coordinate.
In the Nariai case there is no proper center for the spherical symmetry.
Therefore, we shall call the spacetimes with $ G' =0 $ generalized
Nariai metrics. Finally, the other case corresponds to the GNRSS
spaces, already studied in I, which constitute a distinguished
family of the class of
Kerr-Schild metrics.
\section{Geometrical properties of the SSQV}
\label{s-gpssqv}
We denote by $ t $, $ r $, $ \theta $, $ \varphi $ the coordinates of the
forms~\rf{met2} and~\rf{met-gnrss}, since no confusion can arise
in what follows.
\subsection{Generalized Nariai metrics}
\label{ss-gnm}
Using an orthonormal cobasis defined as
\beq
\begin{array}{l}
\th^0 = \Bigl(1-{H \over 2} \Bigr) \, dT - {H \over 2} \, dr, \quad \th^1 =
\Bigl( 1 + {H \over 2} \Bigr) dr + {H \over 2} \, dT, \cr
\th^2 = \gamma \, d \theta, \quad \th^3 = \gamma \, \sin\theta \, d\varphi,
\end{array}
\eeq
we see that the Riemann tensor has as independent components
($ A' \equiv dA/dr $)
\beq
R_{0101} = -{H'' / 2}, \quad R_{2323} = {1 / \gamma^2}.
\eeq
The Ricci tensor is characterized by
\beq
R_{00} = - R_{11} = -{H'' / 2}, \quad R_{22} =  R_{33} = {1 / \gamma^2}.
\eeq
The scalar curvature is simply $ R = H'' + 2/\gamma^2 $, and the Einstein
tensor has the following non-zero components
\beq
G_{00}= - G_{11} = {1 / \gamma^2} \equiv \Lambda_0,
\quad G_{22} = G_{33} = - {H'' / 2},
\eeq
where we have already identified the energy-matter density
with the value of the cosmological constant, thanks to the presence of the
Nariai solution inside this family. The isotropic solution is very important
in order to set the type of spacetime to be chosen for the core
of the object. An immediate calculation yields $ H = \Lambda_0 r^2 + b r + c
$, where $ b $ and $ c $ are arbitrary constants. Without losing generality,
we can set $ b $, $ c  = 0 $ (as they are clearly gauge freedoms for any
spacetime in the family). Thus, the {\it only isotropic} quantum vacuum
belonging to the family is the Nariai solution.
\subsection{The GNRSS metrics}
\label{ss-gnrss2}
We refer the reader to I for details on the geometrical properties
of these spacetimes. The main results: the isotropic GNRSS is the de Sitter
solution, given by $ H = (\Lambda_0/3) r^2 $, and the exterior metric for a
static black hole also belongs to the GNRSS family (e.g. the
Reissner-Nordstr{\"o}m
one corresponds to $ H = 2M/r - Q^2/r^2 $, where $ M $ is the ADM mass and
$ Q $ its electric charge.)

It is worth recalling that both these isotropic quantum
vacuums are regular at the origin.\footnote{The only singular 
``isotropic'' solution belongs to the
GNRSS family and is actually the Schwarzschild solution. It is nevertheless
a true, or classical, vacuum solution.}
\section{The interior structure of a regular black hole}
\label{s-srbh}
In this section, we introduce what may be viewed as a trial model
for a regular black hole. We do not address here the question of
relating each
claimed spacetime solution with a (quantized) source origin. For now,
we will only consider the main features of the desired structure.

For the exterior region, we can think of any (classical-dominated) black 
hole
solution. For instance, the Schwarzschild, Reissner-Nordstr{\"o}m,
Kottler-Trefftz \cite{kramer}--\cite{trefftz}, or the like, solutions.
The third is simply a
Schwarzschild-de Sitter solution which would allow for the introduction of
a non-zero value of the (true) cosmological constant. In fact, in I
we added the quantum corrections to these exteriors coming from vacuum
polarization,
but these contributions are only relevant near the surface of the collapsed
body and can be often dismissed. A remarkable property is that all these 
spacetimes belong to the GNRSS class\footnote{They can be obtained by 
choosing $ H(r) = 2 M/ r ^Ö Q^2/r^2 + \Lambda_1 r^2 $, where $ M $, $ Q $ and 
$ \le $ are the mass, the charge and a external cosmological constant, 
respectively.}  Then, for the interior of the body,
we select a certain SSQV solution representing a transient state, relevant
until the point when the spacetime becomes isotropized at the core of the
object. The exterior
solution and the core are almost fixed. On the contrary, the transient
region is fairly free, as long as a study of the source origin is not
carried out.
The fact that Nariai, or de Sitter, and Schwarzschild solutions cannot be
directly connected, forces one (as mentioned in the Introduction), to
consider
a smooth transition zone. This is dominated by the SSQV solution though,
eventually, complete isotropization is expected to occur deep
inside the body, owing to the dominance of vacuum polarization.
If another type of effects dominate, as for instance strings, then the core
could change. We now turn the attention towards the reliability of the
two previous families as candidates for solving the present scheme.
\subsection{The matching of generalized Nariai and GNRSS spacetimes}
\label{ss-mgng}
Assuming the exterior solution is actually a member of the GNRSS
family, we shall first discuss  if it is possible, for any such member,
to match with some member of the generalized Nariai class.
This will provide a direct check of the possibility for an eventual
combination between the two.

The general form of a hypersurface that clearly adjusts itself to the
spherical symmetry of any of these spacetimes is:
\beq
\Sigma : \cases{\theta = \lambda_{\theta}, \cr
\varphi = \lambda_{\varphi} , \cr
r = r(\lambda) , \cr
t = t(\lambda) ,}
\eeq
where $ \{\lambda,\lambda_{\theta}, \lambda_{\varphi} \} $
are the parameters
of the hypersurface. Its corresponding tangent vectors are
\beq
{\vec e}_{\theta} = \partial_{\lambda_{\theta}} \stackrel{\Sigma}{=}
\partial_{\theta}, \quad {\vec e}_{\varphi} = \partial_{\lambda_{\varphi}}
\stackrel{\Sigma}{=} \partial_{\varphi}, \quad {\vec e}_{\lambda} =
\partial_{\lambda} \stackrel{\Sigma}{=} {\dot r} \partial_{r}
+ {\dot t} \partial_t,
\eeq
where the dot means ``derivative with respect to $ \lambda $''.
The normal one-form is then ($ {\bf n} \cdot {\vec e}_i = 0 $,
$ i = \{\theta, \varphi, \lambda\} $)
\beq
{\bf n} \stackrel{\Sigma}{=} \sigma({\dot r} dt - {\dot t} dr),
\eeq
If $ \bf n $ is a null one-form, $ \sigma $ is a free function. Otherwise,
$ {\bf n} $ can be normalized, e.g. $ {\bf n} \cdot {\bf n} = \pm 1 $, and~$
\sigma $ can be chosen to be
\beq
\sigma_{\pm} \stackrel{\Sigma}{=} {\pm 1 \over \sqrt{|{\dot t}^2-
{\dot r}^2+H ({\dot r} -{\dot t})^2}|}.
\eeq

The first junction conditions reduce to the coincidence of the first
differential form of $ \Sigma $ at each spacetime. One must thus identify
both hypersurfaces in some way. The identification of
$ (\lambda_i)_1 $ with $ (\lambda_i)_2 $ (1 and 2 label each of
the spacetimes) is clearly most natural, due to the symmetry of
the above scheme. This yields (if 1 labels the GNRSS spacetime and 2 the
generalized Nariai one)
\beq
\label{first1-2}
r_1(\lambda) = \gamma = {\rm const.}, \\
\label{first2-2}
-{\dot t_1}^2(1+H_1) = {\dot r_2}^2-{\dot t_2}^2+H_2 ({\dot r_2} + {\dot 
t_2})^2
.
\eeq

The second set of junction conditions comes from
(we do not use any singular mass shell,
albeit this could be added without problem)
\beq
\label{eq-hij}
\Bigl[{\cal H}_{ij}\Bigr] = 0,
\eeq
where $ {\cal H}_{ij} $ is defined by
\beq
{\cal H}_{ij} \stackrel{\Sigma}{\equiv} - m_{\rho}
\biggl({\partial^2 \phi^{\rho}
\over \partial \lambda^i \partial \lambda^j} +
{ \hbox{$ \Gamma $}}^{\rho}_{\mu \nu}
{\partial \phi^{\mu} \over \partial \lambda^i}
{\partial \phi^{\nu} \over \partial \lambda^j} \biggr) .
\eeq
Here $ \vec{m} $ is any vector that completes the set
$ \{ \vec e_{i} \} $ to form a vectorial basis of the
manifold\footnote{For the cases when $ \bf n $
is non-null, $ \vec m $ can be chosen simply
as $ \vec n $, and $ {\cal H}_{ij} $ becomes the second
fundamental form. $ {\cal H}_{ij} $ allows for
dealing with a transition at the event horizon of the black hole.}
(see e.g. \cite{marsseno} and I).
We will select, for the GNRSS spacetime,
$$ \vec m_1 = {\dot r_1} \partial_{t_1} - {\dot t_1} \partial_{r_1} . $$
This choice has the property that $ \vec m_1 \cdot {\bf n}_1 =
\sigma_1 ({\dot r_1}^2 + {\dot t_1}^2) $. It is a convenient one
because, if it vanishes, the
hypersurface $ \Sigma $ becomes degenerate and the
joining process itself cannot be carried out. For the Nariai spacetimes,
as there is no preferred radial coordinate to be identified
with $ r_1 $, we have to leave $ \vec m_2 $ free, and see if there is some
choice that makes the matching possible. Of course, it must at least satisfy
$ {\vec m} \cdot {\bf n}_2 \neq 0 $, if $ \Sigma $ is a hypersurface.
Furthermore,
$ \phi^{\rho} (\lambda) $ are the parametric equations of the
hypersurface ($ \{ \phi^0, \allowbreak\phi^1,\phi^2, \phi^3 \} =
\{t,r,\theta, \varphi\}_{\Sigma}  $), and $ \Gamma^{\rho}_{\mu \nu} $
the connection coefficients.

Eqs.~\rf{eq-hij} yields, for {\em any} $ \vec m_2 $,
\beq
\label{second1-2}
r_1 {\dot t_1} = 0,
\eeq
An immediate consequence of~\rf{second1-2} is that the
matching between any generalized Nariai spacetime and
any member of the GNRSS is {\em not} possible (if
$ \dot t_1 =0 $, since also $ \dot r_1 =0 $, the matching
becomes meaningless, i.e. it is only valid for an ``instant'' and for
a specific ``place''.) The only way to overcome this impossibility is
calling for singular mass shells. However our aim is to avoid such 
unphysical
situation. We do not mean by this that the transition zone cannot become
small in comparison with the isotropic region, but that we disregard exact
singular mass shells as physical solutions for the structure of the 
collapsed body because it would affect even to the claimed smoothness of the 
model (see also the conclusions in~\cite{bp}).

To summarize, if vacuum polarization is to be the dominant quantum effect,
the most simple way to construct a regular black hole is to build
it upon GNRSS spacetimes .
\subsection{The matching of two GNRSS spacetimes}
\label{ss-mtgn}
We refer the reader to I, where explicit expressions for the matching
conditions are given. It follows from them, that the only physically
acceptable solutions
are the hypersurfaces defined by (now 1 and 2 each label a GNRSS space)
$ r_1 =
r_2 = R = {\rm const.} $ The rest of the conditions,
Eqs.~\rf{first2-2},~\rf{eq-hij},
are rather simple, namely,
\beq
H_1|_{r_1=R} = H_2|_{r_2=R}, \quad H'_1|_{r_1=R} =
H'_2|_{r_2=R}, \quad {\dot t_1} = {\dot t_2},
\eeq
where $ H'|_{r=R} \equiv dH(r)/dr|_{r=R} $. This also shows that the
coordinate system chosen in~\rf{met-gnrss} is in fact a privileged one,
in which the metric takes an explicit $ C^1 $ form.

Thus, the most plausible scheme is the one depicted in Tab.~\ref{f-fqios}.
Obviously, the
properties of the intermediate interior region are most important, the
rest already having a clear interpretation in physical terms, including
quantum fields that may act as sources for the core, i.e. de Sitter 
spacetime.

A natural simplification of the scheme above is to demand that the
interior GNRSS solution tends to the de Sitter solution as $ r \to 0 $.
In that case, we get the simplified picture of Tab.~\ref{f-rqibh}.
This is a common assumption to be found in the  literature (see e.g.
\cite{ip}--\cite{ds}, \cite{magli1}--\cite{ayon}). It was also made in I.

We shall now study the case of Tab.~\ref{f-fqios} in detail because it
is free from
the drawback of demanding $ r \to 0 $, i.e. the imposition that the 
spacetime
should be extended to regions where
quantum effects play a dominant role, and where the classical notions of
space and time cease to be valid, together with the existence of an inner
Cauchy horizon (see e.g. \cite{mps}), which
will lie now very far from the scheme depicted in Tab.~\ref{f-fqios}, as we 
will show.
However the second scheme actually allows for a complete, and easier,
study of the semiclassical zone, as shown in paper I.
\begin{table}
$$\begin{array}{||c|c||c|c||}
\hline
{\rm \bf Core} & {\rm \bf Interior} & {\rm\bf  Exterior} & \infty \cr
\hline
\hline
\hbox{de Sitter} &  \hbox{GNRSS} & {\rm Non-Rotating} & {\rm Cosmological} 
\cr
& & \hbox{Black Hole} & \hbox{Term}\cr
\hline
\end{array} $$
\caption{\label{f-fqios} {\small Physically meaningful scheme for a static
 regular  black hole
interior dominated by quantum vacuum polarization. The cosmological term is
optional.}}
\end{table}
\begin{table}
$$
\begin{array}{||c||c|c||}
\hline
{\rm \bf Interior} & {\rm\bf  Exterior} & \infty \cr
\hline
\hline
\hbox{GNRSS} & {\rm Non-Rotating} & {\rm Cosmological} \cr
  & \hbox{Black Hole} & \hbox{Term}\cr
\hline
\end{array}
$$ \caption{\label{f-rqibh}Simplified picture corresponding to
Tab.~\ref{f-fqios}
where $\hbox{GNRSS}_{int} \to \hbox{de Sitter} $, for $ r \to 0 $.}
\end{table}

\subsection{Constraints for the de Sitter-GNRSS model}
\label{ss-cdsg}
Looking at Tab.~\ref{f-fqios} and noticing that the core and the
exterior parts
of the model only depend on $ \Lambda_0 $, $ M $, $ Q $,
it turns out that two sets of constraints for the
unknown function $ H_{int} $ appear. Of course, the parameters of the core,
i.e. $ \Lambda_0 $, and those coming from the exterior solution,
i.e. the mass, the charge, and the (exterior) cosmological constant,
also impose some restrictions, but we will consider them as free data.

The matching of the de Sitter and the interior GNRSS yields
\beq
\label{cond-ds}
\Sigma_{\rm dS}: r_{\rm dS} = r_{int} \equiv R_{\rm dS} = \hbox{const.},
\quad (\Lambda_0 / 3) R^2_{\rm dS} = H_{int}(R_{\rm dS}), \cr
(2 \Lambda_0 / 3) R_{\rm dS} = H'_{int}(R_{\rm dS}),
\eeq
where $ R_{\rm dS} $ may be interpreted as the
``de Sitter radius'' of the object, the scale where isotropization
takes place.
Analogously, the matching of the interior and the exterior GNRSS gives
\beq
\label{cond-bod}
\Sigma_{\rm body}: {\tilde r}_{\rm body} = r_{\rm body} \equiv R = 
\hbox{const.},
\quad H_{int}(R) = H_{ext}(R), \cr
H'_{int}(R) = H'_{ext}(R),
\eeq
where we have set $ {\tilde r} $ for the interior region simply because it 
is
different from $ R_{\rm dS} $ (in fact, only $ R_{\rm dS} < R $ makes 
sense).

We thus have two conditions at $ R_{\rm dS} $ and two more at $ R $.
For the scheme~\ref{f-rqibh} we have two conditions at $ r \to 0 $ and two
more at $ R $. Both schemes are then similar,
but now we have a new unknown, namely, $ R_{\rm dS} $. In order to
see the changes, it is worth solving a particular set of models. We will
consider the analogous of the two-power models of I. In those,
$ H_{int} (r) = a_{p} r^{p} + a_q r^{q} $, where $ p $, $ q $
are real numbers (obviously $ p < q $ would suffice). It is worth noticing
that the case $ p = q $ is incompatible with the
conditions~\rf{cond-ds},~\rf{cond-bod}, e.g. the de Sitter spacetime alone
is not sufficient to fulfill the internal structure of the black hole.

For the exterior solution we shall choose a
``Schwarzschild-de Sitter'' model which accounts for current
astrophysical-cosmological observations on black holes (see e.g.
\cite{astro}--\cite{ries}),
whereas for the core we use a de Sitter solution. The functions $ H $ are,
respectively, $ H_{ext} = 2 m /r + ( \le / 3 ) r^2 $, and
$ H_{\rm core} = ( \li  / 3 ) r^2 $, where $ \le $, $ \li $ are
the cosmological terms of the exterior region and the core,
respectively (one expects $ \le << \li $). Isolating the
coefficients $ a_p $, $ a_q $ from Eqs.~\rf{cond-ds},~\rf{cond-bod}, using 
the
previous expressions for $ H_{ext} $ and
$ H_{int} $, we get (remember $ p \neq q $)
\beq
a_p = \biggl({q-2 \over q-p}\biggr)
\biggl({ \li \over 3 R_{\rm dS}^{p-2}}\biggr) =
\biggl({1 + q \over q-p}\biggr)  \biggl({ 2 m \over R^{q + 1}}\biggr)
- \biggl({2 - q \over q - p }\biggr) \biggl({ \le \over 3 R^{p-2}}\biggr) ,
\eeq
and an analogous result for $ a_q $, interchanging $ p $ and $ q $.
If $ q = 2 $,
one gets immediately $ m= 0 $. Therefore $ p, q \neq 2 $.
We do not recover now the lowest power model of paper I, because
the body suddenly becomes isotropized at $ r = R_{\rm dS} $. Moreover,
if $ p = -1 $, necessarily $ q = -1 $. Therefore $ p = q $ and there is
no solution of~\rf{cond-ds},~\rf{cond-bod}. The constraints are
then ($ p \neq q $, $ p, q \neq -1, 2$)
\beq
\label{eq-r}
{\li \over 3} = \biggl( { q + 1 \over q - 2  } \biggr)
\biggl( { 2 m \over R R^2_{\rm dS}} \biggr)
\biggl( {R_{\rm dS} \over R} \biggr)^p
+  {\le \over 3 } \biggl( {R_{\rm dS} \over R} \biggr)^{p-2} \cr
= \biggl( { p + 1 \over p - 2  } \biggr)
\biggl( { 2 m \over R R^2_{\rm dS}} \biggr)
\biggl( {R_{\rm dS} \over R} \biggr)^q
+  {\le \over 3 } \biggl( {R_{\rm dS} \over R}\biggr)^{q-2}.
\eeq
From the expressions above, we realize that $ R_{\rm dS} $ and $ R $
can be obtained in terms of $ m $, $ \le $, $ \li $, $ p $ and $ q $.
However, in order to arrive to explicit algebraical relations, we will
disregard the contribution of $ \le $. Another reason to do so comes
from observational arguments. We expect $ \li >> \le $. Furthermore,
we also expect $ R_{dS} < R $, $ R \leq 2m $ (the collapsed object is at,
or beyond, the event horizon) so that $ \le R^2 << 1 $ and the contribution
of the terms with  $ \le $ in~\rf{eq-r} are completely negligible in front 
of
the rest. We then get
\beq
\label{eq-rm}
R = \biggl( { 6 m \over R^2_{\rm dS} \li} \biggr)
\biggl[ \biggl({p-2 \over p+1}\biggr)^p
\biggl({q+1 \over q-2}\biggr)^q\biggr]^{1 \over q -p}.
\eeq
If we define (as in I, $ \le $ neglected),
$ R_Q \equiv \sqrt[3]{6m / \li} $, we can rewrite~\rf{eq-rm} as
\beq
\label{eq-rq}
R = \biggl( {R^3_Q \over R^2_{\rm dS} } \biggr)
\biggl[ \biggl({p-2 \over p+1}\biggr)^p
\biggl({q+1 \over q-2}\biggr)^q\biggr]^{1 \over q -p}.
\eeq
Setting $ R \equiv 10^{\beta} R_Q $,
$ R_{\rm dS} \equiv 10^{-\gamma} R_Q $, we get
\beq
\beta = { (q-2) \log |(q+1)/(q-2)| + (p-2) \log|(p-2)/(p+1)|
\over 3(q-p)}, \\
\gamma = { (q+1) \log|(q-2)/(q+1)| + (p+1) \log|(p+1)/(p-2)|
\over 3 ( q - p ) }.
\eeq
One readily sees that $ \beta + \gamma > 0 $ for most situations. It is
unnecessary to carry out a complete study of the exponents. Our
aim is just to show the main features of the internal black hole structure
by means
of this simple (but already non-trivial) model. Some results are
displayed in Table~\ref{t-1}, for a number of different exponents.
It is shown there that, regardless of the powers,  the results
are very similar, and equivalent to those obtained in paper I. Thus with
high confidence the whole picture emerges as a likely structure for
static black holes.
One should also notice that, here, the problems of ascribing a de Sitter 
core
when  $ r \to 0 $ is solved, although, numerically, $ R $ does
scarcely change with
respect to the results of the simplified model in I.
The proximity of the de Sitter region to
the limiting hypersurface of the body has also been claimed as convenient
in \cite{burinskii},
where the Kerr black hole is studied, and justifies the idea of effective
mass shells, though only in an {\it effective} sense, i.e.
not arriving to the creation of a singular distribution.

Finally let us add that $ R_Q $ is obtained when one tries to match
the de Sitter
core with an exterior, Schwarzschild black hole metric, imposing only the
natural condition $ H_{\rm dS} (R) = H_{\rm Schw}(R) $.
We already knew that this direct junction was impossible, unless mass shells
are introduced. However, it is noticeable that the relevant order of
magnitude is in complete agreement with that of the rigorous result.
\begin{table}
$$
\begin{array}{||c||c|c|c|c|c|c||}
\hline\hline
& ( 1000, 3 )  & (3 , 100 ) & (3 , 4) & ( -3,-2 ) & (-100,-4) &
(-5000,-1000) \cr
\hline
\hline
R_{\rm dS} & 0.999 & 0.99 & 0.4 & 0.86 & 0.996  & 1 - 10^{-7} \cr
\hline
R & 1.001 & 1.01 & 0.9 & 1.36 & 1.003 & 1 +10^{-7}  \cr
\hline
\end{array} $$
\caption{\label{t-1} {\small Values for $ R_{\rm dS} $, $ R $ 
($ R_{\rm dS} < R $) in
units of $ R _Q $ for several powers. In all cases, the
values are very close to each other. Moreover, $ R_Q $ is very far away from
the scale of regularizaton for any astrophysical, or more massive, object.
Therefore they all lie in the semiclassical regime.}}
\end{table}
\section{Extension to other possibilities for interiors}
\label{s-eip}
In this section, we turn our attention to other possible candidates for the
interior of a black hole. In the previous ones, the models were built upon
the idea that vacuum polarization or supergravity domain wall 
\cite{burinskii}--\cite{cvs} would play an essential role.
But it is also
interesting to check whether a stringy black hole could actually describe
the core of a black hole
or, even, its transitory region.
There are many results that point towards a correspondence between
semiclassical black holes and stringy ones
\cite{sv}--\cite{damour}, \cite{burinskii}. Among
the most widely studied solutions there are the supersymmetric ones.
However, for our
aim of considering a self-gravitating black hole, an approximate
solution will suffice, provided the transition takes place at the
semiclassical
level. As with the previous models, this fact can be eventually justified
by carrying out, {\it  a posteriori}, a numerical analysis of the solutions
obtained,  (see paper I).
We just impose here the most elementary conditions that are able to
produce a stringy
interior in the black hole, and we will leave to further work the 
calculation
of the self-gravitating string (this issue being still under study,
see e.g. \cite{damven}).

Stringy black holes do not correspond to SSQV solutions in general.
Thus, we need
to perform their matching with a GNRSS solution. However it seems
more natural to carry out such a calculation in general.
Namely, to find the conditions for two static spherically
symmetric spacetimes to match with each other. In fact, this is a
well-known program, specially if one does not include the possibility
of null hypersurfaces, see e. g. \cite{israel2,robson}.
For the case of null hypersurfaces one can use the formalism
in \cite{cd}, allowing a transition at a horizon.
The addition of null hypersurfaces could be
of interest in those
attempts at studying radiative processes in the interior region
of the black hole, \cite{balpoi}--\cite{israel1}. However, it would be
a nuisance to have to
alternate all the time between two different pictures, depending
on which process we would like to study. It is possible and more interesting
to perform a general study which can take into account different
behaviors of the matter and radiation in the black hole interior. Thus, we
will consider a general hypersurface which may turn out to be null at
any time or spatial point of the interior region. The
necessary formalism
to deal with this situation can be found in \cite{marsseno}.
The general solution to this question is to be found in~\cite{mps}.
However the representation of the spacetimes being used is
not a common one for the spacetime metrics we shall deal with,
and the translation of those results to each of our cases
turns out to be even more tedious than a direct computation.
Thus, we shall follow here a different approach. It is important
to note that, once the conditions are obtained, the process
of matching  itself
becomes irrelevant. Therefore, we shall try to obtain the conditions
as directly as possible.
\section{The matching of static spherically symmetric \allowbreak
spacetimes}
\label{s-mssss}
Our aim here is to match two spacetimes that share the existence of an
integrable Killing field and spherical symmetry. In order to get the
most natural junction, we need to take profit of both symmetries
exhaustively. The spherical symmetry is easy to identify in both
spacetimes. The metric can always be written, for any of them, as
\beq
\label{met-g}
ds^2 = g_{AB}(R) \, dx^A \, dx^B + G^2(R) \, d\Omega^2,
\eeq
where $ A $, $ B = T, R $; $ \partial_T $ has been chosen to be
the integrable Killing vector and $ d\Omega^2 = d \theta^2
+ \sin^2 \theta \, d\varphi^2 $. Moreover, if $ G'(R) = 0 $, we know
that the spacetimes belong to the generalized Nariai family, in which case
they only match with another member of its own family, as
is easy to show. Thus, we will only deal with the situation
$ G'(R) \neq 0 $. In this case, a direct redefinition of the
$ R $ coordinate allows us to write
\beq
\label{met-r}
ds^2 = g_{AB}(r) \, dx^A \, dx^B + r^2 \, d\Omega^2,
\eeq
where $ A $, $ B = T, r $.

Spherical symmetry has thus been completely used. We now extract
consequences from the presence of  $ \partial_T $, an integrable Killing
field.
The natural thing to do is to identify both vector fields, i.e.
$ \partial_{T_1} \stackrel{\Sigma}{\equiv} \partial_{T_2} $.
However this is not a right choice, in general. If a Killing vector
is multiplied by a constant factor, the resulting vector field is
obviously a Killing vector field. Therefore, normalizing each
Killing vector, when possible, gives the natural way to identify
them. This is implemented in the junction process,
if the hypersurfaces are spacelike or timelike everywhere, where the
calculations are easier as well (see the previous footnote). On the
contrary, in the general case, we cannot rely on such normalization,
and $ \vec m $ takes the role of $ {\bf  n} $.
But $ \vec m $ is an extrinsic object with respect to the hypersurface.
Thus, one has to take care of the identification process.
The results do not depend on $ \vec m $, provided the
identification is the one desired. In Sect.~\ref{s-srbh} this was
readily implemented. Let us show how it can be done now.

Any metric of interest (to our purpose) can be written as
(recall the coordinate change to obtain~\rf{eq-fn0},
setting now $ F = 1- H $)
\beq
ds^2 = -(1-H) \, dT^2 + {2H \over g} \, dT \, dr
+ {1+H \over g^2} \, dr^2 + r^2 \, d\Omega^2,
\eeq
where $ H $, $ g \neq 0 $ are functions of $ r $ only.
Looking back to expressions~\rf{met-3},~\rf{met-gnrss}
and to their
equivalence,  we will put
now $ dT = g_0 dt + ( g_0 - g_0^{-1} ) dr $, where $ g_0 $ is a constant,
that will be related with the function $ g $, as we shall see in a moment.
With this coordinate change the metric takes the form\footnote{It is also
equivalent to leave the $ T $ coordinate unchanged, and change $ \vec m $.
However, we prefer the characterization above, because it
directly tells us which is the  $ t $ coordinate to be identified in
both spacetimes.}
\beq
\label{met-l}
ds^2 = - g_0^2(1-H) \, dt^2 + 2 {\tilde G} \, dt \, dr
+ {\tilde F} \, dr^2 + r^2 \, d\Omega^2,
\eeq
where $ {\tilde G} = g_0^2 ( H - 1) +1 + H (g_0 - g )/ g $, and
$ {\tilde F} = 2 + g_0^2 ( H - 1 ) + (g_0 - g)\allowbreak
[ 2 H / g + 2 / g_0^2 g + ( g_0-g )( H + 1 )/ g_0^2 g ] $.

The junction conditions are then
\beq
[ r ] = 0, \quad [{\dot t} ] = 0  \\
\label{cond-3}
\bigl[{\tilde H}\bigr] {\dot t}^2 + 2 \bigl[{\tilde G}\bigr] {\dot t}{\dot 
r}
+ {\tilde F} {\dot r}^2 = 0, \\
\label{cond-4}
[{\tilde F}] {\dot t}{\ddot r} - [{\tilde G}] ({\ddot t}{\dot t}
- {\ddot r}{\dot r}) + [{\tilde H}] {\dot r}{\ddot t}
+ [{\tilde G}' ] {\dot r}^3 + [ {\tilde H}' -({\tilde F}'/2) ]
{\dot r}^2 {\dot t} - [{\tilde H}' ] {\dot t}^3 / 2 = 0,
\eeq
where $ [ f ] \stackrel{\Sigma}{=} f_2 - f_1 $, and where
we have put $ {\tilde H} \equiv g_0^2( H - 1 ) + 1 $.
In~\rf{cond-3} and~\rf{cond-4} $ \dot t $ and $ \dot r $
are either $ \dot t_1 $, $ \dot r_1 $ or $ \dot t_2 $, $ \dot r_2 $,
and $ A' \equiv dA(r)/dr|_{r=r(\lambda)} $.
The same conditions lead, in general, to a second order ordinary 
differential
equation for $ r $.
In principle there is the possibility for asymptotic
stopping solutions, i.e. solutions for which
$ r \to {\rm const.} $ as $ t \to \infty $, and also for null ones.
A complete study of these possibilities is worth pursuing, after
having decided which are the specific spacetimes to be considered
in accordance with our purposes, i.e. provided
some knowledge about $ H(r) $, $ g(r) $ is available. A special case
is of great interest, since it constitutes the solution towards
which any transitory solution should converge, in order to have a
regular black hole interior, namely $ r_1 = r_2 = R = {\rm const.} $.
Under this restriction, the conditions become, simply,
\beq
\bigl[{\dot t} \bigr] = 0, \quad
\bigl[{\tilde H} \bigr] = 0,
\eeq
and
\beq
2 \bigl[{\tilde G} \bigr] {\ddot t} - \bigl[{\tilde H}']  {\dot t}^2 = 0,
\eeq
where $ t $ is either $ t_1 $ or $ t_2 $. Choosing $ g_0 $ as
$ g_{\Sigma} $ one gets $ [ {\tilde G} ] = 0 $
(the same result comes out directly in the case when the normal
vector of $ \Sigma $ is non-null). The last conditions
become then $ [{\tilde H}' ] = 0 $.
Thus, we arrive at the conclusion that the conditions emerging
from the matching of two spherically symmetric spacetimes
with an integrable Killing vector field are, for the case
$ r = R = {\rm const.} $ and taking
the maximum identification between them,
\beq
[ \tilde H ] = 0 , \quad [{\tilde H}' ] = 0,
\eeq
where $ {\tilde H}  \equiv g_{\Sigma}^2 ( H - 1 ) + 1 $. An intrinsic
characterization, valid for any representation of the form~\rf{met-g}
or~\rf{met-r} (those are expressions that are often dealt with)
is $ {\tilde H} \equiv -g_{\Sigma} ({\vec \xi} \cdot {\vec \xi}) + 1 $,
$ g_{\Sigma} \equiv [G' / |\det (g_{AB})|^{1/2}]_{r=R} $,
where $ {\vec \xi} $ is the Killing vector associated with the staticity
of the solution (in some regions) of~\rf{met-g} or~\rf{met-r}.
It is not difficult to realize that the first condition on $ {\tilde H} $ is
nothing but the requirement of the mass function to be continuous across the
hypersurface, while the second one is related with the continuity of the
radial stress,
or pressure, see e.g.~\rf{eq-p}.
\section{An application to supersymmetric stringy black \allowbreak holes}
\label{s-assbh}
The semiclassical expressions for supersymmetric stringy black holes are
well-established (see e.g. \cite{horowitz,horpol} and references therein).
There are also other objects of
interest, such as black strings, higher dimensional black holes, etc.
In all cases one looks for a correspondence principle with general
relativistic black holes. This equivalence, or transition, is usually
reflected in the strength of the coupling constant, or the entropy (see e.g.
\cite{horowitz}--\cite{damour} and references therein).
Here we take a different viewpoint, coming from
the above scheme, which turns out to be valid because all the fields
are in the semiclassical domain. Thus, it would be interesting to see
whether both approaches complement each other, or which are the
restrictions that may be induced from the current scheme.
The most interesting case is that of a self-gravitating string
(see e.g. \cite{damven,damour}). However the necessary ingredients
---specially the corresponding spacetime metric--- in order to tackle
this problem are still under study. Here we will
deal with the most simple (and widely considered) case only, i.e.
that of a supersymmetric back hole.

A family of such black holes, related with electrically charged
black holes, is given by (see \cite{horowitz,horpol} for details)
\beq
ds^2 = - f^{-1/2}(r) \Bigl( 1 - {r_0 \over r} \Bigr)
\, dt^2 + f^{1/2}(r) \biggl[ \Bigl( 1 - {r_0 \over r} \Bigr)^{-1}
\, dr^2 + r^2 \, d\Omega^2 \biggr],
\eeq
where
$ f(r) = \prod_{i=1}^{4} [ 1 + (r_0 \sinh^2 \alpha_i/r) ] $, and where
the $ \alpha_i $ are related with the integer charges of the D-branes
being used. If the correspondence occurs at a constant value of $ r $, we 
get
\beq
r_1 + r_0 \sinh^2 \alpha = r_2 f_2^{1/4} (r_2) \equiv R =
{\rm const.}\\
{2m \over R} - {Q^2 \over R^2} = 1
+ \biggl[\biggl({r_0 \over r} - 1\biggr)
\biggl( 1 + {r f' \over 4 f} \biggr)^2 \biggr]_{\Sigma_2}, \\
- {2m \over R^2} + {2 Q^2 \over R^3} =
\biggl\{ \biggl( 1 + {r f' \over 4 f} \biggr)^2
\biggl[{f' \over 2 f}\biggl(1- {r_0 \over r} \biggr) -
2 {r_0 \over r^2} \biggr] \biggr\}_{\Sigma_2},
\eeq
where we have used $ g_{\Sigma} = G'|_{r = G^{-1}(R)} $,
$ G(r) = r f^{1/4}(r) $, and $ {\vec \xi} \cdot {\vec \xi}
= - f^{-1/2} ( 1 - r/r_0) $. The subscript $ \Sigma_2 $ means that all these
quantities refer to the interior region, to be evaluated at $ r =r_2 $. For
the exterior metric, we have put $ \alpha_i = \alpha_j \equiv \alpha $,
for all $ i $, $ j $, because the exterior metric is that of a
Reissner-Nordstr{\"o}m black hole, for which
\beq
\begin{array}{l}
2m = r_0 \cosh 2 \alpha, \quad Q^2 = r_0^2 \sinh^2 \alpha
\, \cosh^2 \alpha, \cr
r_0 = 2 \sqrt{m^2 - Q^2}, \quad 2 \sinh^2 \alpha = -1
+ m/\!\sqrt{m^2 - Q^2 },
\end{array}
\eeq
where $ m $ is the (ADM) mass of the black hole and $ Q $ is
its electric change.
Since
$ f_2(r_2) = \prod_{i=1}^4[1 + (r_0 \sinh^2 \alpha_i/r)]_{\Sigma_2} $,
the above conditions yield $ R $ as a function of six of the seven
parameters,
$ M, Q, (r_0)_2, \alpha_i $.
Detailed analysis shows that these conditions are easily
fulfilled when $ r_0 \to 0 $, $ \alpha_i \to \pm \infty $, with
$ r_0 \sinh^2 \alpha_i $ fixed. The resulting $ R $ is very close
to $ R_0 \equiv m + \sqrt{m^2-Q^2} $, i.e. the event horizon of the
black hole.
It is interesting to notice that $ r f^{1/4}(r) $ is the radial coordinate,
which has a direct interpretation in terms of the ``size'' of the object,
and not of $ r $ alone. All this being in complete agreement with the
expected transitions for extreme, and nearly extreme, supersymmetric
black holes. The same idea could be extended to self-gravitating strings,
where the physical scheme becomes more interesting for the higher
plausibility of having regular interiors in this case. For instance,
the expected
order of magnitude of $ R $, \cite{damven}, should be recovered. This
issue will be the  matter of subsequent research.
\section{Conclusions}
In this work we have investigated, under quite general conditions, the
question of using Einstein's theory
of gravitation ---extended to include semiclassical
effects--- with the purpose to
constraint the physical structure of the emerging
spacetime solutions that might be suitable
for the description of the interiors
of non-rotating black holes.

In the first part of the work we have made extensive
use of general ideas, coming from plausible, expected quantum
contributions to the energy-momentum tensor, in order to get
the main features of the
resulting models in terms of a spacetime viewpoint.
For instance, we have exploited the idea that vacuum
Polarization or supergravity, e.g. domain wall, may indeed play
an essential role in the interior region. We have obtained
the result that only two families
fulfill the imposed requirement and, moreover, we have shown
that {\it only one} of them
is suitable for representing black hole interiors, what
is certainly a most remarkable
result. Moreover, we have extended
the models of paper I in order to solve the problem of demanding
{\it isotropization} near the regularization scale and have proven that this
is indeed possible, with little changes in the previous results.

Then we have turned our attention to stringy black holes. Since the
solutions we are interested in ---self-gravitating strings--- still need to 
be
studied in more
detail, we have just started this program by first giving
the general conditions to be fulfilled by any spacetime
with spherical symmetry and with some static region. Finally, we
have applied the results obtained  to a supersymmetric black hole, as
a preliminar case.
The result is that the proposed models are indeed generically compatible,
specially concerning the extreme limit. This last situation is
precisely the same for
which the correspondence between semiclassical black holes and stringy
ones has been recently confirmed in the literature
(see e.g. \cite{damven,damour}).

Our overall conclusion  is the following.
In the first place, the results in paper
I have opened a new window for the search
of a compatible quantum field that, once regularized, may
yield the same result for, at least, a particular energy-momentum tensor
inside the general family of models considered. Second, once a corresponding
Einsteinian metric associated with a quantum model is known,
the scheme developed here might be certainly well suited to
check the consistency of the involved physical parameters and even, in some
cases, to assign explicit values to them.
All these results, as a whole,
compel us to believe that black hole singularities are
likely to be removed by quantum effects, at least in the non-rotating case.
\footnote{The rotating case, which is of
major astrophysical interest, and the rotating and electrically charged one,
which may be associated with spinning particles, seem to yield
results very similar to the ones
presented here, see e.g. \cite{burinskii}. This
is also the outcome of preliminary calculations
of ours, to be reported elsewhere \cite{behm}.}
\section*{Acknowledgements}
The authors acknowledge valuable discussions with A. Burinskii
(and also his careful reading of the manuscript) and with G.
Magli. This work has been supported by CICYT (Spain), project
BFM2000-0810 and by CIRIT (Generalitat de Catalunya), contracts
1999SGR-00257 and 2000ACES 00017.

\end{document}